\documentstyle[12pt,graphicx]{article}

\newcommand{\bwide}{\begin{widetext}}
\newcommand{\ewide}{\end{widetext}}
\newcommand{\beq}[1]{\begin{equation} \label{(#1)}}
\newcommand{\eeq}{\end{equation}}
\newcommand{\bea}[1]{\begin{eqnarray} \label{(#1)}}
\newcommand{\eea}{\end{eqnarray}}

\newcommand{\ba}{\begin{array}}
\newcommand{\ea}{\end{array}}
\def\lsim{\mathrel{\vcenter{\hbox{$<$}\nointerlineskip\hbox{$\sim$}}}}
\def\gsim{\mathrel{\vcenter{\hbox{$>$}\nointerlineskip\hbox{$\sim$}}}}
\newcommand{\Egzk}{E_{\rm GZK}}
\newcommand{\Eres}{E_{\rm res}}
\begin{document}

\title{%
\vskip-1.5in\vskip-6pt \hfill 
{\rm\normalsize UCLA/04/TEP/8} \\
\vskip-0.2in~\\
Bounds on Relic Neutrino Masses in the Z-burst Model
}

\author{Graciela Gelmini \\
  {\it Department of Physics and Astronomy, UCLA} \\
  {\it Los Angeles, CA 90095-1547, USA} \\
  {\tt gelmini@physics.ucla.edu} \\
  \and
  Gabriele Varieschi \\
  {\it Department of Physics, Loyola Marymount University} \\
  {\it One LMU Dr., Los Angeles, CA  90045-2659, USA} \\
  {\tt gvariesc@lmu.edu}
  \and
  Thomas Weiler \\
  {\it Department of Physics, Vanderbilt University} \\
  {\it Nashville, TN 37235-1807, USA}; and \\
  {\it Theory Division, CERN, CH-1211, Geneva 23, Switzerland} \\
  {\tt tom.weiler@vanderbilt.edu} }

\maketitle

\begin{abstract}
Neutrinos from far-away sources annihilating at the Z resonance on relic 
neutrinos  may give origin to the extreme-energy cosmic rays (EECR).
If ``Z-bursts'' are responsible for the EECR events, then 
we show that the non-observation of cosmic
ray events at energies above 2 $\times 10^{20}$ eV by the AGASA Collaboration 
implies a lower bound $\sim 0.3$~eV on the relic neutrino mass.
Since this mass exceeds the mass-squared differences inferred from
oscillation physics, the bound in fact applies to all three neutrino masses.
Together with the upper bound provided by comparisons of the CMB anisotropy
with large-scale structure, this bound 
leaves only a small interval for neutrino masses around 0.3 eV,
if Z-bursts are to explain the existing EECR events.

\end{abstract}

\newpage
\section{Introduction}
The existence of extreme-energy cosmic rays (EECR) with energies above the
Greisen-Zatsepin-Kuzmin (GZK) cutoff~\cite{gzk}  of about $5\times 10^{19}$ eV,
presents an outstanding problem~\cite{data}.  Nucleons and photons with those
energies have short attenuation lengths and could only come from distances of
100 Mpc or less~\cite{50Mpc,40Mpc}, while plausible
astrophysical sources for those energetic particles are much farther away.

Data from the HiRes experiment \cite{HiRes} brought the violation 
of the GZK cutoff into question.  
Yet the EE events in even the HiRes data set remain unexplained since the 
local Universe ($\sim 100$~Mpc) is devoid of strong candidate sources.
This controversy will be solved conclusively by the Pierre Auger 
hybrid observatory~\cite{Auger}, perhaps as soon as summer of 2005.
Here we assume that the published AGASA spectrum is correct.
Interestingly, it is the absence of events above $2\times 10^{20}$~eV in
these data, rather than the presence of events above 
$E_{\rm GZK}\sim 5\times 10^{19}$~eV, 
that motivates this work.

Among the solutions proposed for the origin of the highest energy events
 observed by AGASA, an elegant and economical solution  to this problem 
is the ``Z-burst'' mechanism: annihilation at the $Z$-resonance of 
extreme-energy neutrinos ($\nu_{\rm EE}$) coming from remote sources, 
and relic background neutrinos within $\sim 50$~Mpc of Earth, 
produces  the nucleon and photon EECRs~\cite{Zbursts}.
The observed EECRs from the Z-bursts are the emission analogues 
of the absorption features (Z-dips) predicted long ago~\cite{Zdips}.
One of the most appealing features of the $Z$-burst mechanism is that the
energy scale of $\gsim 10^{20}$~eV at which the
unexpected events have been detected, is generated naturally.
The $Z$-resonance occurs when the energy of the incoming $\nu_{\rm EE}$ is 
\begin{equation}
\Eres = \frac{M_Z^2}{2~m_\nu} 
	= 4\times 10^{21}{\rm eV}\,\left(\frac{\rm eV}{m_\nu}\right)\,,
\label{Eres}
\end{equation}
where $m_\nu$ is any of the three masses of the relic neutrinos. 
Given the lower limit $\sim 0.04$~eV deduced from atmospheric oscillations for 
the heaviest neutrino mass, at least one $\Eres$ is below $10^{23}$~eV.

Since the individual energies of the 
nucleons and photons emerging from the Z-burst cannot exceed the total 
energy of the burst,
$\Eres$ is the new end-point of the EECR energy-spectrum in this 
mechanism. 
Partitioning this burst energy among the $\langle N\rangle \sim 40$ 
final state particles, one arrives at precisely the primary energies needed 
to produce events observed above the GZK energy.
 
Combining their recent  measurements of the
anisotropies of the Cosmic Microwave Background
(CMB) radiation with data on the large scale structure
of the Universe, WMAP~\cite{WMAP} produced a strong limit on neutrino masses, 
$\Sigma_i m_{\nu_i} < 0.69$~eV.
Since a single neutrino mass above $\sim 0.04$~eV implies
near mass-degeneracy for all three active neutrinos
(given the mass-squared splittings from neutrino oscillation data,
described below)
one has $m_\nu < 0.23$~eV at the 95\% CL. 
However, objections~\cite{objections} to the priors assumed, 
or to the data sets included,
have led to a relaxed bound $\Sigma_i m_{\nu_i} < 1$~eV, or $m_\nu < 0.33$~eV.
A subsequent analysis~\cite{BMT} which includes WMAP, 
2dF, and SDSS data, and another which includes these and galaxy 
cluster data~\cite{ASB},
have arrived at the bound, $\Sigma_i m_{\nu_i} \lsim 0.7$~eV.
Still another analysis~\cite{CLP} with CMB and LSS data finds 
$\Sigma_i m_{\nu_i} \lsim 1$~eV, but finds $\Sigma_i m_{\nu_i} \lsim 0.6$~eV
when priors from supernova data and the Hubble Key Project are included.
These newer bounds are very similar to the original WMAP bound.

In this work we focus on a particular feature of the AGASA spectrum,
namely the end-point energy.  Requiring that the Z-burst mechanism not 
overproduce 
events beyond this AGASA end-point, we derive a lower bound on the neutrino 
mass.
The significance of an end-point energy for constraining model fluxes has
 been noted
much earlier in the context of topological defect decay~\cite{PJ-PS}.
A more ambitious project would be to actually 
fit the neutrino mass to the full AGASA spectrum.
This has been done, by Fodor, Katz and Ringwald (FKR)~\cite{FKR}.
However, the allowed range of the neutrino mass that results is not tight.
Rather, it depends sensitively on how one parameterizes 
the transition from non-burst spectrum to burst spectrum 
near and above the ankle.
FKR find $\pm 2\sigma$ fitted mass ranges of 0.1 to 1.3~eV 
if all EE events are assumed to originate with Z-bursts,
0.02 to 0.8~eV if an additional extra-galactic source of EE protons is allowed,
and 1 to 7~eV if an additional Galactic-halo source of EE protons is allowed.

Super-Kamiokande has provided a strong evidence for the oscillation in
atmospheric showers of two neutrino species with  
mass-squared splitting $\delta m^2 $ = $m_3^2-m_2^2$ = $(1-3)
 \times 10^{-3}$~eV,
consisting of nearly equal amounts of $\nu_\mu$ and 
$\nu_\tau$, and little or no $\nu_e$~\cite{SK}.  
If neutrino masses are hierarchical, like the other leptons and quarks,
then $\sqrt{\delta m^2} \simeq 0.04\,{\rm eV} \equiv m_{\rm SK}$
 \underline{\sl is} the mass of
the heavier of the two oscillating neutrinos (call it $\nu_{\rm SK}$).
Some previous works~\cite{gk1,gv} on Z-burst production of the EECRs 
have made this assumption.
However we will show in this work 
that $m_\nu\sim 0.04$~eV is not compatible with the AGASA end-point spectrum 
if Z-bursts produce the EECR.
The reason is that with $m_\nu = 0.04$~eV, the new EECR cutoff 
$\Eres \simeq 10^{23}$~eV
predicts too many super GZK EECR events beyond the AGASA end-point.

As the relic neutrino mass increases, $\Eres\propto 1/m_\nu$ decreases,
and the features in the EE spectrum move progressively to lower energies.  
In particular, the number of events
predicted at high energy decreases, and the number beyond the AGASA end-point 
becomes compatible with zero for $m_\nu \gsim0.3$~eV.
Note that our result is much sharper than, but compatible with, 
the allowed mass ranges of the FKR models with no additional EE proton source, 
or with an additional extra-galactic source of EE protons.

In the next section we present our simulations and the resulting 
spectrum of EECR in detail.

\section{Spectrum of EECR from Z-bursts}

We have performed simulations of the photon, nucleon and
neutrino fluxes coming from a uniform distribution of ``Z-bursts",
namely  $\nu \bar{\nu}$ annihilations at the Z pole ($\nu
\bar{\nu} \rightarrow Z \rightarrow p \gamma ...$).  The burst energy 
is given in Eq.~1.
Our simulations cover the range $m_{\rm SK}\le m_\nu < 1$~eV
for relic neutrino mass.  
The Z-bursts were simulated using PYTHIA
6.125~\cite{PYTHIA}, and the absorption of photons and nucleons was modeled 
using energy-attenuation
lengths provided by Bhattacharjee and Sigl~\cite{BS}, supplemented by 
radio-background models by Protheroe and Biermann~\cite{PB}.

We simulated a uniform distribution of about $10^7$ Z
particles up to a maximum $z_{max}=2$. 
If other Z-burst spatial distributions are used,
the main features of the spectrum of the ultra-high energy
nucleons, photons and neutrinos remain the same as given below.

The decay of the Z bosons through all possible channels was done automatically
by the PYTHIA routines~\cite{PYTHIA}, using the default options of this
program. For comparison with our later figures, 
we show in Fig.~1 the spectra 
given by PYTHIA, normalized per Z boson, for $m_\nu =0.3$~eV;
redshifts and energy absorption are not included in PYTHIA.  
The multiplicities that PYTHIA gives per Z-decay are 
1.6 nucleons, 17 photons, 15 $\nu_e$, 30 $\nu_{\mu}$ and 0.23 $\nu_{\tau}$
(in each case counting particles and antiparticles).

\begin{figure*}
\vspace*{-35.0mm}
\hspace*{-25.0mm}
\includegraphics{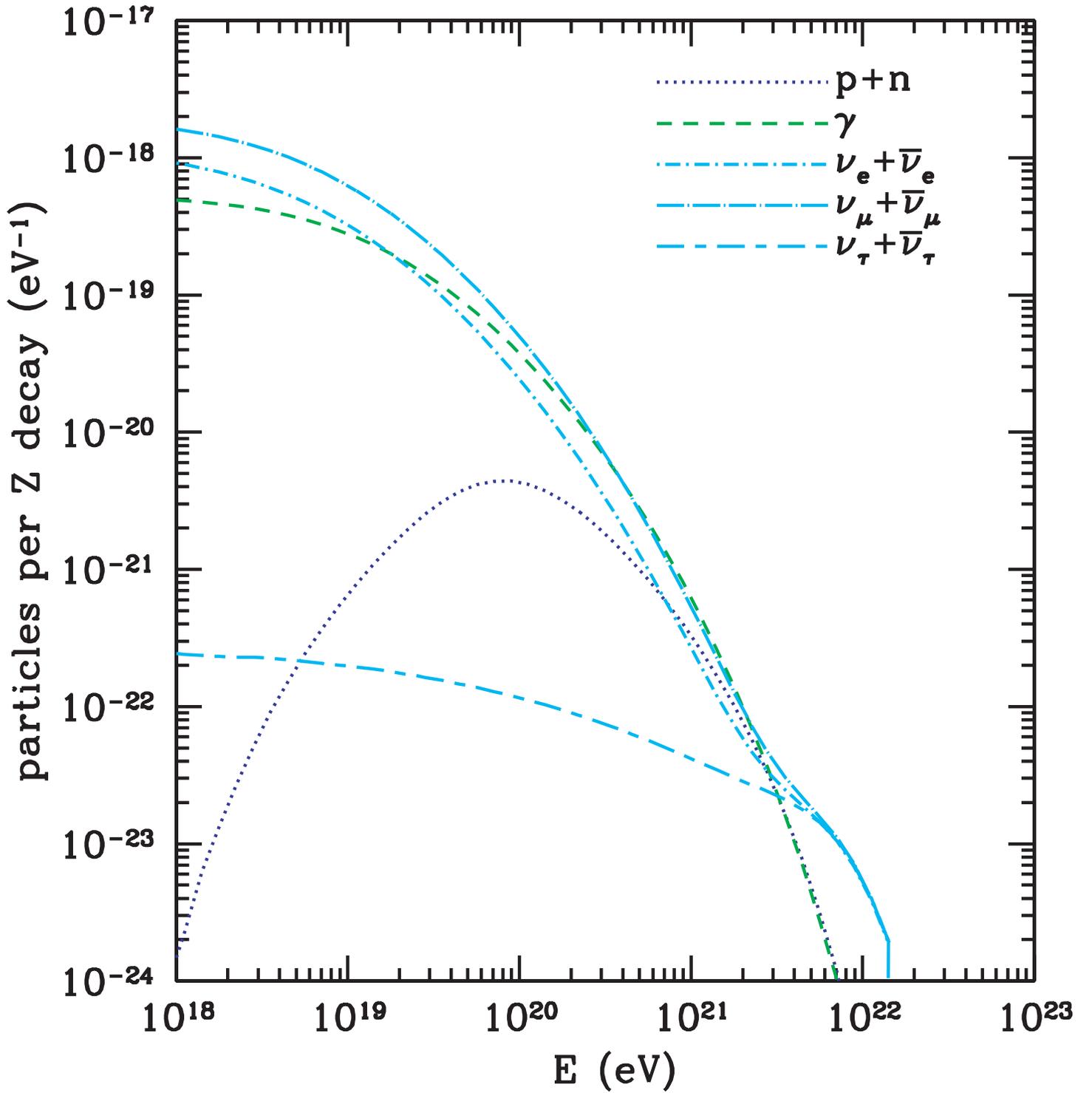}
\caption{\label{fig1}
Spectra of stable particles produced per Z-decay by PYTHIA 
(no absorption or redshift included).
The resonant energy in this example is $E_{\rm res}=1.3\times 10^{22}$~eV,
coming from the choice $m_\nu=0.3$~eV.}
\end{figure*}

In our simulation, each Z boson generated by PYTHIA was placed on the ``event
list'' of the cascade generator at a randomly selected distance.  The cascade 
of decay products was then boosted. The
$\gamma$ factor corresponding to an energy  $E_\nu=E_{\rm res}$ in Eq. (1) is
\begin{equation}
\gamma\equiv{\frac{E_\nu+m_\nu}{M_z}} \simeq \frac{M_z}{2\,m_\nu}
	\simeq 0.46 \,\left(\frac{0.1\,{\rm eV}}{m_\nu}\right)
\times 10^{12}\,.   
\label{eq:sim3}
\end{equation}  
We then followed the propagation of the nucleons, photons and neutrinos 
resulting from the boosted Z-decays. 
The gamma factor of the secondary particles was corrected to include their 
red-shifting subsequent to the Z-decay.

Each neutrino, nucleon or
photon was created by PYTHIA in the
initial cascade at a fixed position, with  fixed energy and direction of
motion with respect to the Earth frame of reference.
The distance  each particle had to travel before reaching  Earth was
compared with the appropriate attenuation length in space for the
particle energy.  If the distance  was smaller than the
attenuation length, the particle was assumed to  reach  Earth unchanged.
In the opposite case, the energy and
momentum of the particle were degraded by a factor $\frac{1}{e}$ 
after traveling
a distance equal to the attenuation length (and the particle was
placed again in the list constituting the cascade at the new 
position,  with the degraded energy and momentum).

Neutrinos  do not interact in their propagation. 
Thus, the energy spectra of
the three kinds of neutrinos were simply generated by summing the number of
particles over energy bins and normalizing to neutrino multiplicities
 per Z times 
the total number of Z particles used.
On the other hand, nucleons and photons do suffer energy-absorption
 interactions,
and here we included energy absorption.
The propagation process for nucleons and photons was continued
until particles reached Earth and were counted in the final spectra, 
or until particle energies became too small to be significant.  
In the latter case, the particles were simply discarded from the cascade.
At this point, the final nucleon and photon spectra were appropriately 
normalized to the total number of Z-particles used,
as was done with the neutrinos. 
The results are given in Fig.~2 for $m_\nu=0.3$,
with a fit to the six most energetic occupied bins in 
the AGASA data~\cite{AGASAdata}.
These bins consist of 24 events spanning the energy range
from $10^{19.8}$~eV to $10^{20.3}$~eV.

\begin{figure*}
\vspace*{-35.0mm}
\hspace*{-25.0mm}
\includegraphics{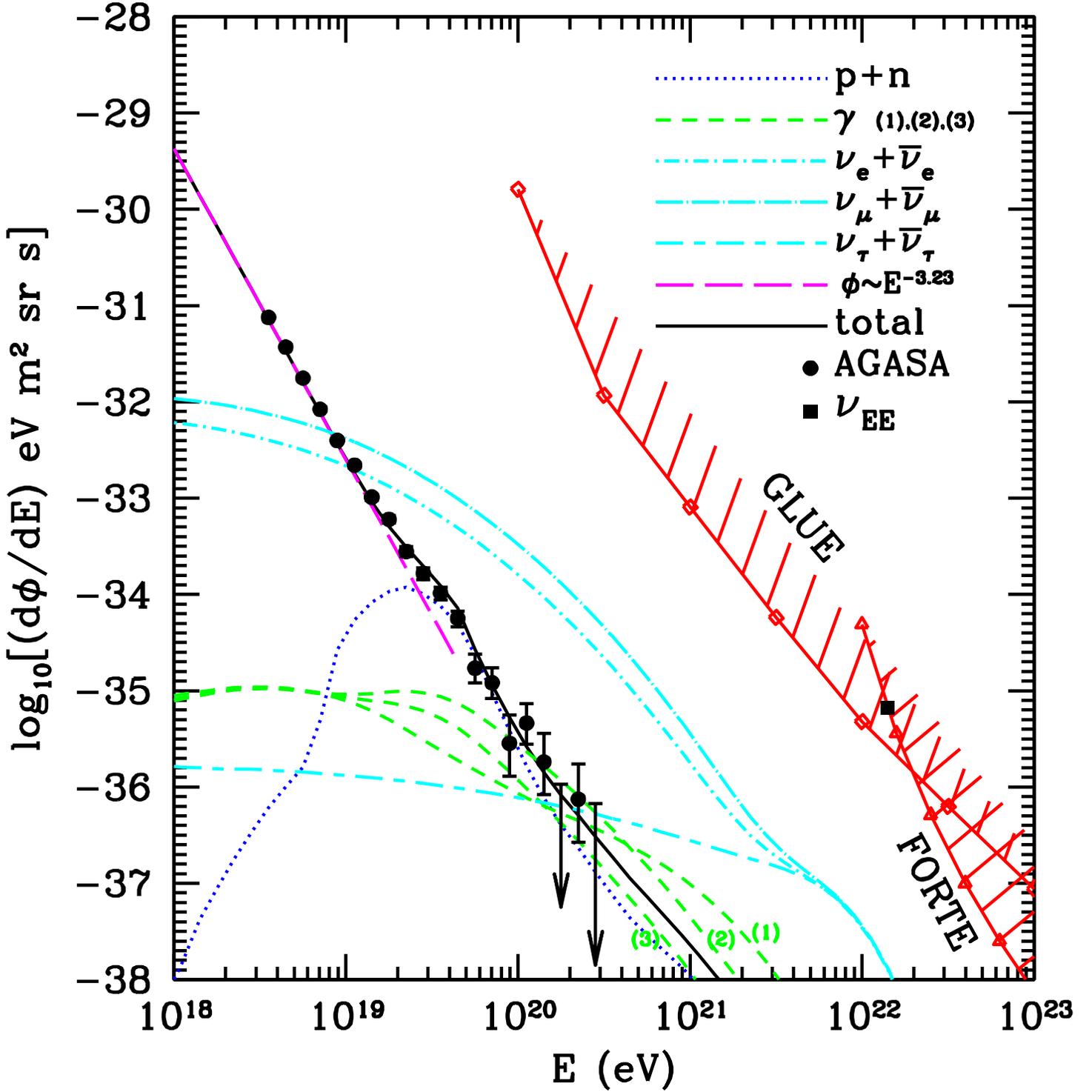}
\caption{\label{fig2}
Predicted spectra for $m_\nu=$0.3 eV from Z-bursts with a uniform 
distribution 
up to $z=2$, added to a power law spectrum which fits the
data below the ankle~$\sim 10^{19}$~eV, and terminates at 
$4\times 10^{19}$~eV. 
It is seen that 
EECR primaries above the ankle are nearly 100\% nucleons up to 
$10^{20}$~eV, and  photons plus nucleons at higher energies.
Also shown is the EE neutrino flux at the unique resonance energy 
which produces the required Z-burst rate.}
\end{figure*}

\begin{figure*}
\vspace*{-35.0mm}
\hspace*{-25.0mm}
\includegraphics{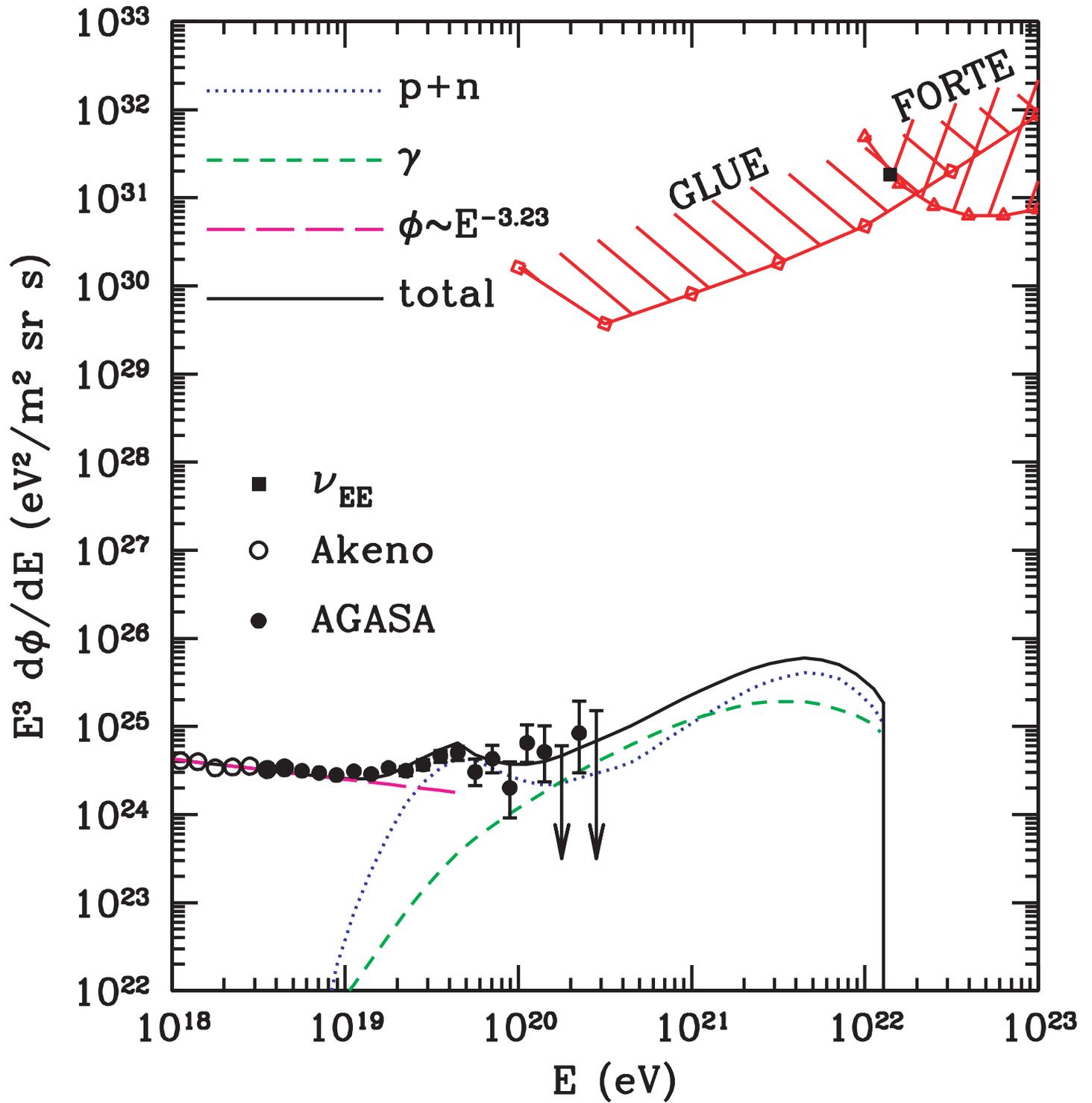}
\caption{\label{fig3}
As in Fig. 2 but with the flux multiplied by $E^3$.
The photon flux shown is the lowest of the three of 
Fig.~2 (labeled as (3)).}
\end{figure*}

The energy-attenuation of the nucleons and especially photons 
requires more discussion.
For nucleon attenuation, uncertainties are small, and so the
predicted GZK suppression is highly credible.
For the nucleon energy attenuation, we used the length 
given by Bhattacharjee and Sigl
in the  Fig.~9 of Ref.~\cite{{BS}}, 
based on results from Ref.~\cite{stecker} and~\cite{halzen}.
However, for photon energy-attenuation,
the length is quite uncertain, 
due to the uncertain spectrum of the absorbing radio background. 
The photon flux shown in Fig.~2  as curve~(1) results from 
using the  attenuation length shown in 
Fig.~11  of Ref.\ \cite{BS},  based on  observations of
Clark et al.~\cite{clark}.  
Protheroe and Biermann~\cite{PB} produced two models for the radio background
which lead to shorter interaction lengths (and therefore more absorption) 
than those based on Clark et al.
From the interaction lengths they provided, we constructed approximate 
attenuation lengths for the models of Protheroe and Biermann.
We did so by reducing the attenuation length based on Clark et al. 
by the ratio of the respective interaction lengths, and obtained 
the curves (2) and (3) in Fig.~2. 
Since mean interaction lengths and mean energy-attenuation lengths do not
have exactly the same energy dependence,
our construction is an approximation. 
However,  we believe the three curves 
which we display give a good representation of the 
possible range of predicted
photon fluxes.

In this work, we seek a bound on the \underline{\sl excess} 
of events predicted.
Therefore, we adopt the lowest observable photon flux, curve number (3), 
associated with the maximum photon absorption, 
when computing the total event number. This is the photon flux shown in Fig.~3.
Notice that with this choice, the photon flux does not become comparable to 
the nucleon flux until $E\sim 2\times 10^{20}$~eV.
We add to the Z-burst proton and photon fluxes 
a power law spectrum with slope $-3.23$,
terminated somewhat arbitrarily (as it is not known
where this component actually dies out) at $4\times 10^{19}$~eV.
Such a power law was found by AGASA to fit the
data below the ankle, from $4\times 10^{17}$ to $10^{19}$ eV
(see Table V of \cite{naganowatson}).

\section{Lower bound on the relic neutrino mass}
In Fig.~4 the observed and predicted 
integrated number of 
EECR event rates above $7\times 10^{19}$~eV are shown for 
different values of the relic neutrino mass.
The Z-burst fluxes have been normalized to reproduce the 24 
highest-energy events observed by AGASA ~\cite{AGASAdata} 
in the energy range from $10^{19.8}$~eV to $10^{20.4}$~eV.
These events are spread among the first six bins of Fig.~4.

The area under each curve at the final bin shows the
integrated number of events from $2.5\times 10^{20}$~eV to infinity
predicted by that so-normalized Z-burst flux. 
This number of excess events can also be read off from the value of the curve 
at $E=10^{19.8}$~eV less the 24 normalizing events.
Since AGASA has observed no events above 2.5$\times 10^{20}$~eV
although their aperture remains constant with increasing 
energy~\cite{AGASAdata},
it is justified to call the events predicted beyond 2.5$\times 10^{20}$~eV
an excess; we label the predicted number of excess events as $N_{\rm ex}$.
In Table~1 the predicted event excess above $2.5\times 10^{20}$~eV
is shown as a function of the neutrino mass.  
Also shown is the Poisson probability that a fluctuation downward from the 
predicted value gives zero events, as observed.
This latter quantity is just $e^{-N_{\rm ex}}$.

\begin{table}
\begin{tabular}{||c|c|c||} \hline\hline
neutrino mass & predicted event excess $N_{\rm ex}$ & 
Poisson probability $e^{-N_{\rm ex}}$ \\ \hline\hline
0.6 eV		& 2.29 		& 10.1\%        \\ \hline
0.5 eV		& 2.76		& 6.3\%		\\ \hline
0.4 eV		& 3.43		& 3.2\%		\\ \hline
0.3 eV		& 4.43		& 1.2\%		\\ \hline
0.2 eV		& 6.09		& 0.23\%	\\ \hline
0.1 eV 		& 9.67		& 0.0063\%	\\ \hline\hline
\end{tabular}
\caption{Poisson probabilities that no events are observed above 
the AGASA end-point,
given the excess predicted by the Z-burst mechanism as a function
 of neutrino mass.}
\end{table}

The probabilities show that masses lighter than $\sim 0.3$~eV
have a probability less than a percent.  Expressed as Gaussian variances,
one infers from the percentages in the table that 
$m_\nu\gsim 0.2$~eV is disfavored at $\sim 3\sigma$ while  
$m_\nu> 0.4$~eV is disfavored at only $\sim 2\sigma$.
Together with the cosmological limit on the neutrino mass discussed earlier,
the value $m_\nu\sim 0.3$~eV emerges as less than robust, but viable.

We remark that our definition of ``excess'' as events above 
$2.5\times 10^{20}$~eV
is somewhat conservative in that the two highest energy AGASA events
occur at $E\sim 2\times 10^{20}$~eV ($\pm 25\%$ experimental 
energy-resolution),
nearer to the beginning of the highest-energy occupied bin than to its end.
On the other hand, we have not considered in our analysis the 
very-highest energy event, at $3\times 10^{20}$~eV, from the 
Fly's Eye experiment.  This is in the spirit of neglecting fluorescent data,
e.g. HiRes,
as we have chosen to focus in a self-consistent way on just the 
ground-scintillator data set from AGASA.

\begin{figure*}
\vspace*{-35.0mm}
\hspace*{-25.0mm}
\includegraphics{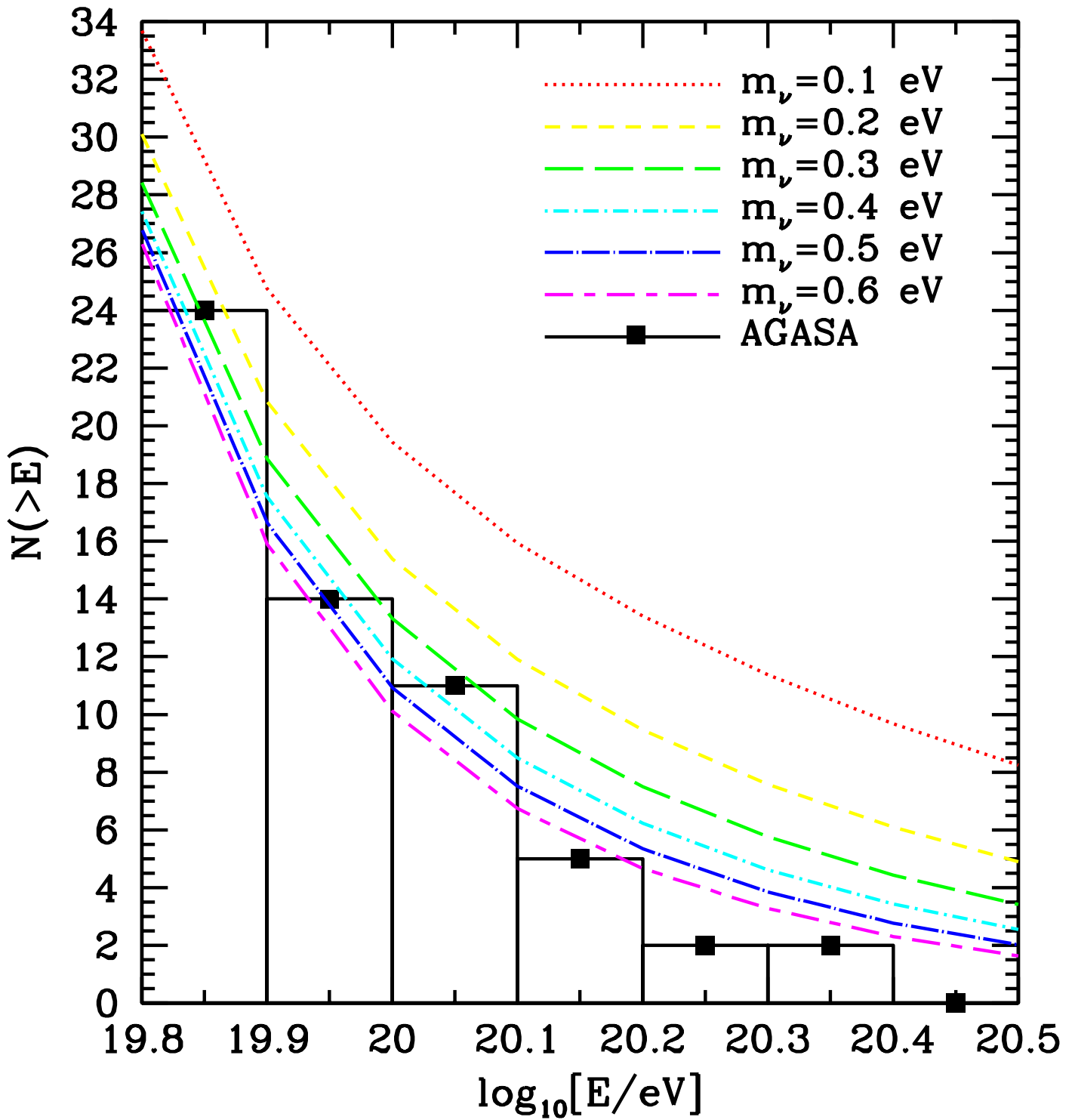}
\caption{\label{fig4} Integrated Z-burst fluxes for different
relic neutrino masses, normalized to reproduce the 24
 highest-energy AGASA events 
in the energy range $10^{19.8}$ to $10^{20.4}$~eV 
(corresponding to the first six of seven bins shown here).
The neutrino mass parameterizes the curves, with lighter mass
 at top and heavier
mass at bottom.}
\end{figure*}

There are three immediate inferences to be drawn from the 
preferred neutrino mass value. 
The first is that 
since the energy of no secondary from the Z-burst may exceed the burst energy,
the model predicts a new cutoff for EECRs.
With 0.3 eV relic neutrinos,
the cutoff energy is $E_{\rm res} \simeq 1.3 \times 10^{22}$~eV.
Given the large mean-multiplicity in Z-decay, $\langle N\rangle\sim 40$,
a more realistic cutoff in the model is $\sim 10^{21}$~eV.
Future experiments such as Auger, ANITA, RICE, EUSO, and 
SALSA may be able to test this prediction.
The second inference is that with $m_\nu\sim 0.3$~eV,
a significant clustering of relic neutrinos around very large 
galactic super-clusters is expected \cite{Singh:2002de}.  
This clustering could be evidenced 
as an enhancement of the EECR flux coming from the direction of M87, 
just 16~Mpc away, near the Virgo center. 
And the third inference is that given the smallness of mass-squared 
differences 
inferred from neutrino oscillation experiments,
$m_\nu\sim 0.3$~eV implies near mass-degeneracy for the three active neutrinos.
Mass-degeneracy enhances the possibility of directly observing absorptive 
``Z-dips'' in the EE neutrino spectrum~\cite{Zdips}.

\section{Discussion}
\label{sec:discuss}
We note that our analysis of the spectral shape of the AGASA data
does not depend on the issue of rate.
In particular, it does not depend on the EECR neutrino flux
and it does not depend on the value of the relic neutrino target density.
The latter could,  in principle, 
be enhanced by either gravitational clustering 
or by a lepton asymmetry.  In practice, simulations argue against 
gravitational clustering for light ($\lsim 1$~eV) neutrinos on
 all but the very largest
super-Galactic mass-scales \cite{Singh:2002de} (which comprise
 a small solid angle of sky),
and BBN physics argues against a lepton asymmetry enhancement above
 $\sim 20$\% \cite{lepasym}.
(Strictly speaking, the BBN limit is strongest for $\nu_e$'s, but the
large lepton mixing angles inferred from oscillation studies 
extend the limit equally 
to $\nu_\mu$'s and $\nu_\tau$'s.)
Regardless of these kinds of rate arguments, our bound on the 
neutrino mass holds.

Farrar and Piran~\cite{farrarpiran} have argued 
that any mechanism accounting for the events beyond the GZK cutoff
should also account for the events down to the ankle, including the 
observed isotropy and spectral smoothness. 
It is seen in our Fig.~2 that the Z-burst 
mechanism can accommodate the position of the ankle and all the events
 above it, 
if the  position of the ankle is close to that measured by AGASA, 
at $E_{\rm ankle} = 10^{19.0}$~eV 
(see \cite{naganowatson}, in particular Table V, and references therein).
Fly's Eye claims a value for $E_{\rm ankle}\sim 10^{18.5}$~eV, 
a factor of three smaller than the AGASA value 
(see Table V of \cite{naganowatson}).
This value is difficult to accommodate with the Z-burst mechanism,
although one could perhaps gain agreement by increasing $z_{max}$ of the 
astrophysical neutrino sources.
This would allow for more red-shifting of the nucleons,
thereby lowering the soft end of the Z-burst spectrum.
A higher $z_{max}$ may be expected, for example, 
if the source is decaying massive particles or topological defects.

The fit of the AGASA data with our total flux provides the normalization of
the photon and nucleon differential fluxes $F$, denoted as $d\phi/dE$ in 
Figs.~2 and 3 (in this case
$F_{\rm AGASA}$ = 
$6.6 \times 10^{-15}~({\rm m^2~sr~s})^{-1}$ $F_{\rm PYTHIA}$).
This allows us to determine
the (assumed homogeneous and isotropic) flux of extreme-energy neutrinos
close to the $Z$-resonance energy 
(more precisely, within the energy interval from 
$\Eres/ (1 + z_{max})= \Eres/3$ to $\Eres$) to be
\begin{equation}
F_{\nu_{{\rm EE}}} \simeq 6.6\times 10^{-36} 
\frac{1}{{\rm eV~ m}^2 {\rm ~sr~s}}~~ .  \label{Fnu}
\end{equation}
This flux is shown in Figs.~2 and 3 , with the label ``$\nu_{\rm EE}$". 
Within the level of accuracy of our simulation, 
we only determine the order of magnitude of this flux. 
Nevertheless, the flux requirement is formidable. 
This very-high energy flux 
is at present marginally ruled-out by present GLUE~\cite{GLN} and
 FORTE~\cite{FORTE} bounds,
as can be seen in Figs.~2 and 3~(see also \cite{semikoz}) 
The large required flux in the Z-burst hypothesis is also a challenge
 to theory.
It appears that new physics is required to produce such a large neutrino flux 
at such extreme energy \cite{nuflux}.

In this work, we have not included any effects of extragalactic magnetic 
fields.
For the analysis of this paper to hold as is, 
these fields should be sufficiently small, $< 10^{-9}~G$.
A way to weaken the bound derived here would be to assume the existence
of large enough extragalactic magnetic fields, say $10^{-8}~G$ or larger,
so that through electron pair-production and subsequent synchrotron radiation, 
photons would be eliminated as EECR primaries. 
This would reduce the event count above the AGASA end-point,
and so allow a smaller neutrino mass. However in this case the flux of
EE neutrinos at the Z-resonance energy, $F_{\nu_{\rm EE}}$,
required to produce the observed EECR's would be larger.
Moreover, the resonant energy, $\Eres \propto 1/m_\nu$,
would be larger.  Both the increased flux and the increased energy bring this 
possibility into conflict with the 
GLUE and  FORTE bounds, on EE neutrino fluxes,
shown in Figs.~2 and 3.
Specifically, we find that without the photon contribution to the Z-burst flux,
we would require 
$F_{\nu_{{\rm EE}}} = 7.23\times 10^{-36}\,({\rm eV~m^2~sr~s})^{-1}$ 
at $\Eres=2.08\times 10^{22}~eV$ for $m_\nu=0.2$~eV, 
or $F_{\nu_{{\rm EE}}} = 4.26\times 10^{-36}\,({\rm eV~m^2~sr~s})^{-1}$ 
at $\Eres=4.16\times 10^{22}~eV$  for $m_\nu=0.1$~eV.
Such possibilities are rejected by the GLUE and FORTE bounds.

Another prediction of the Z-burst mechanism is the 
enhancement of the event rate at the GZK cutoff energy, 
$\Egzk\sim 5\times 10^{19}$~eV, 
due to the accumulation of nucleons after energy-attenuation.
Such an enhancement (or ``GZK bump'') is expected from any model
which provides a nucleon spectrum flatter than $E^{-2}$ above $\Egzk$
(spectra falling faster than $E^{-2}$ do not provide sufficient flux 
above $\Egzk$ to make an observable pile-up.).
A hint for such a bump is seen in the AGASA data, Fig.~2 and 3.
However, one must be careful in interpreting the shape as an enhancement.
An alternative explanation of the shape feature has been provided 
in \cite{depression},
namely that another energy-loss mechanism, $p+\gamma\rightarrow p+e^++e^-$,
depletes the flux at energies just below $\Egzk$.

Finally, a  reliable prediction
of the model is that most primaries above the ankle should 
be nucleons  up to $10^{20.0}$~eV or more, and
photons and protons at higher energies.
A much larger event sample is needed to test this prediction.

\section{Conclusions}
The non-observation  of cosmic
ray events at energies higher than $\sim 2\times 10^{20}$ eV by AGASA
provides a lower bound of about 0.3~eV on relic neutrino masses, if 
Z-bursts are responsible for the EECRs. 
This bound together with
comparisons of the microwave background anisotropy 
to today's large-scale structure,
leave a small interval for the relic neutrino mass, centered around 0.3 eV,
if Z-bursts explain the EECRs. 
The $\sim 0.3$~eV mass provides a very accessible signal for 
neutrinoless double beta-decay experiments (assuming neutrinos are Majorana particles), 
and is in fact compatible with a reported positive signal~\cite{Klapdor}.
This $\sim 0.3$~eV mass interval may also be probed by the future KATRIN 
tritium end-point experiment~\cite{katrin},
and by expected improvements to the astrophysical bound 
from weak lensing of distant galaxies by intervening matter 
distributions~\cite{AD02},
and from mining future Planck measurements of the CMB~\cite{PlanckCMB}. 
We note that our small allowed range for $m_\nu$ is incompatible with an
upper bound of 0.15~eV that arises from invoking the see-saw mass 
spectrum 
as a source of successful early-Universe baryogenesis~\cite{leptogenesis}.

We have not included in this work any possible effects of extragalactic 
magnetic fields.
For the analysis of this paper to hold as is, 
these fields should be sufficiently small, $\lsim 10^{-9}~G$.
A survey of the literature shows that 
extragalactic fields much larger than $10^{-9}~G$ are not expected, except
in regions around galaxy clusters.
In any event,  large extragalactic magnetic fields 
which would reduce or eliminate 
the EE photon component of the Z-bursts (thus lowering the neutrino mass
we obtain) would also lead to larger required  EE neutrino fluxes  
at the Z-resonance energy.
This possibility is rejected by the GLUE and FORTE bound.

We have shown that $Z$-bursts  may account  not
only for the super GZK events, but for the ``ankle" 
and all EECR events above it. In doing so, the $Z$-burst mechanism 
meets the ``naturalness" requirements of isotropy and spectral smoothness. 
In our simulation we found
the ``ankle"  close to $E_{ankle} = 10^{19.0}$~eV as measured by AGASA.\ 
We found that most EECR primaries above the ankle
should be nucleons up to about $10^{20.0}$~eV and  
nucleons and photons  at higher energies. 
We also found that primary nucleons do accumulate at the GZK
cutoff energy, which could  account for the slight accumulation seen 
in the data (see, however, \cite{depression}). 

With $m_\nu\sim 0.3$~eV needed to accommodate the AGASA end-point,
the Z-burst hypothesis predicts a new absolute energy cutoff for EECRs, 
at $\Eres \simeq 1.3\times 10^{22}~eV$.
However, due to the large multiplicity in Z-decay, 
the vast majority of incident primaries are expected to
carry only a few percent of this cutoff energy.

In summary, we believe that the $Z$-burst mechanism provides a plausible  
explanation to the puzzle of extreme-energy cosmic rays, not
only for the super GZK events, but for the ``ankle'' and all events above it.
But the hypothesis will stand or fall on the required nearness 
of $m_\nu$ to 0.3~eV, and/or on the proximity of the required 
neutrino flux at $\Eres=1.3\times 10^{22}$~eV to existing flux limits. 

If Z-bursts are invalidated as the source of the observed EECRs,
then we would encourage future, larger experiments (e.g., Auger, EUSO) 
to continue the search for 
Z-burst signatures, for they provide the only known window 
to discovery of the relic neutrino background.  
The Z-burst mechanism is entirely Standard Model particle physics coupled with 
Standard Hot Big-Bang Cosmology.  Consequently, 
the mechanism itself cannot be ruled out.
However, the Z-burst rate 
depends linearly on the resonant-energy neutrino 
flux which Nature provides.  Nature may not be generous.

\vspace{5 mm}

This work was supported in part by NASA grant NAG5-13399,
the US Department of Energy grants DE-FG03-91ER40662, Task C,  
and DE-FG05-85ER40226, and Vanderbilt University's 
Discovery Grant and sabbatical programs.
G.V. was also supported by an award from the Research Corporation. 
We thank A. Kusenko, S. Nussinov and D. Semikoz for many
valuable discussions, P. Biermann and E. Roulet for comments and
suggestions, and the Aspen Center for Physics for a supportive environment 
during part of this work.

\end{document}